\documentclass[aps,pre,reprint,showpacs,amsmath,amssymb,floatfix]{revtex4-1}
\usepackage{graphicx}
\usepackage{bm}
\usepackage{amsmath}  

\begin{document}


\title{Effects of Landau-Lifshitz-Gilbert damping on domain growth}

\author{Kazue Kudo}
\affiliation{Department of Computer Science, Ochanomizu University,
2-1-1 Ohtsuka, Bunkyo-ku, Tokyo 112-8610, Japan}

\date{\today}

\begin{abstract}
 Domain patterns are simulated by the Landau-Lifshitz-Gilbert (LLG)
 equation with an easy-axis anisotropy. 
 If the Gilbert damping is removed from the LLG equation, it merely 
 describes the 
 precession of magnetization with a ferromagnetic interaction.
 However, even without the damping, domains
 that look similar to those of scalar fields are formed,
 and they grow with time. 
 It is demonstrated that the damping has no significant effects on domain
 growth laws and large-scale domain structure.  
 In contrast, small-scale domain structure is affected by the damping.
 The difference in small-scale structure arises from energy dissipation
 due to the damping.
\end{abstract}


\maketitle

\section{Introduction}

Coarsening or phase-ordering dynamics is observed in a wide variety of
systems.  
When a system is quenched from a disordered phase to an ordered phase,
many small domains are formed, and they grow with time.
For example, in the case of an Ising ferromagnet, up-spin and down-spin 
domains are formed, and the characteristic length scale increases
with time. 
The Ising spins can be interpreted as two different kinds of
atoms in the case of a binary alloy.
At the late stage of domain growth in these systems, 
characteristic length $L(t)$
follows a power-law growth law,
\begin{equation}
 L(t)\sim t^n,
\end{equation}
where $n$ is the growth exponent.
The growth laws in scalar fields have been derived by several 
groups: 
$n=1/2$ for non-conserved scalar fields, 
and $n=1/3$ for conserved scalar fields~\cite{Bray1994,
Lifshitz1961,Wagner1961,Ohta1982,
Huse1986,Bray1989,Bray1990,BrayRutenberg1994}.

Similar coarsening dynamics and domain growth have been observed
also in Bose-Einstein condensates (BECs).  
The characteristic length grows as $L(t)\sim t^{2/3}$ in two-dimensional
(2D) binary BECs and ferromagnetic BECs with an easy-axis
anisotropy~\cite{Kudo2013,Hofmann2014,Williamson2016}. 
The same growth exponent $n=2/3$ is found in classical binary fluids in
the inertial hydrodynamic regime~\cite{Bray1994,Furukawa1985}. 
It is remarkable that the same growth law is found in both quantum and
classical systems.
It should be also noted that domain formation and coarsening in BECs occur
even without energy dissipation.
The dynamics in a ferromagnetic BEC can be described not only by the
so-called Gross-Pitaevskii equation, which is a nonlinear Schr\"odinger
equation, 
but also approximately by a modified Landau-Lifshitz equation in which the
interaction between superfluid flow and local magnetization is
incorporated~\cite{Lamacraft2008,Stamper2013,Kawaguchi2012}.
If energy dissipation exists, the equation changes to
an extended Landau-Lifshitz-Gilbert (LLG)
equation~\cite{Kudo2011,Kudo2013,Kawaguchi2012}.  
The normal LLG equation 
is usually used to describe spin dynamics in a ferromagnet.
The LLG equation includes a damping term which is called the Gilbert
damping. When the system has an easy-axis anisotropy, the damping
has the effect to direct a spin to the easy-axis direction.
The Gilbert damping in the LLG equation corresponds to energy
dissipation in a BEC.
In other words, domain formation without energy dissipation in a BEC
implies that domains can be formed without the damping in a ferromagnet.
However, the LLG equation without the damping 
describes merely the precession of magnetization with a ferromagnetic
interaction.

In this paper, we focus on what effects
the damping has on domain formation and domain growth.
Using the LLG equation (without flow terms), we investigate
the magnetic domain growth in a 2D system with an easy-axis anisotropy.
Since our system is simpler than a BEC, we can also give simpler
discussions on what causes domain formation. 
When the easy axis is perpendicular to the $x$-$y$ plane, the system
is an Ising-like ferromagnetic film, and domains in which 
the $z$ component of each spin has almost the same value are formed.
In order to observe domain formation both in damping and no-damping
cases, we limit the initial condition to almost uniform in-plane spins.
Actually, without the damping, domain formation does not occur from an
initial configuration of spins with totally random directions. 
Without the damping, the $z$ component is conserved.
The damping breaks the conservation of the $z$ component as well as energy.
Here, we should note that the growth laws for conserved and
nonconserved scalar fields cannot simply be applied to the no-damping
and damping cases, respectively, in our system.
Although the $z$ component corresponds to the order parameter of a
scalar field, our system has the other two components. 
It is uncertain whether 
the difference in the number of degrees of freedom can be neglected 
in domain formation. 

The rest of the paper is organized as follows.
In Sec.~\ref{sec:model}, we describe the model and numerical
procedures. Energies and the characteristic length scale
are also introduced in this section. 
Results of numerical simulations are shown in Sec.~\ref{sec:rslt}.
Domain patterns at different times and the time evolution of energies
and the average domain size are demonstrated.
Scaling behavior is confirmed in 
correlation functions and structure factors at late times. 
In Sec.~\ref{sec:disc}, we discuss why domain formation can occur even
in the no-damping case, focusing on an almost uniform initial condition.  
Finally, conclusions are given in Sec.~\ref{sec:conc}.

\section{\label{sec:model}Model and Method}

The model we use in numerical simulations is the LLG equation, which is
widely used to describe the spin dynamics in ferromagnets.
The dimensionless normalized form of the LLG equation is written as
\begin{align}
 \frac{\partial\bm{m}}{\partial t} = -\bm{m}\times\bm{h}_{\rm eff}
+ \alpha\bm{m}\times\frac{\partial\bm{m}}{\partial t},
\label{eq:LLG.0}
\end{align}
where $\bm{m}$ is the unit vector of spin, $\alpha$
is the dimensionless Gilbert damping parameter.
We here consider the 2D system lying in the $x$-$y$ plane, and assume
that the system has a uniaxial anisotropy in the $z$ direction and that
no long-range interaction exists.
Then, the dimensionless effective field is given by
\begin{align}
 \bm{h}_{\rm eff} = \nabla^2\bm{m} + C_{\rm ani} m_z\hat{\bm z},
\label{eq:h_eff}
\end{align}
where $C_{\rm ani}$ is the anisotropy parameter, and $\hat{\bm z}$ is the unit
vector in the $z$ direction. 

Equation~\eqref{eq:LLG.0} is mathematically equivalent to
\begin{align}
  \frac{\partial\bm{m}}{\partial t} = 
- \frac{1}{1+\alpha^2}\bm{m}\times\bm{h}_{\rm eff}
+ \frac{\alpha}{1+\alpha^2}\bm{m}\times(\bm{m}\times\bm{h}_{\rm eff}).
\label{eq:LLG.1}
\end{align}
In numerical simulations, we use a Crank-Nicolson method 
to solve Eq.~\eqref{eq:LLG.1}.
The initial condition is given as spins that are aligned in
the $x$ direction with a little random noises: 
$m_x\simeq 1$ and $m_y\simeq m_z \simeq 0$.  
Simulations are performed in the $512\times 512$ lattice with periodic
boundary conditions.
Averages are taken over $20$ independent runs.

The energy in this system is written as
\begin{align}
 E &= E_{\rm int} + E_{\rm ani}
\nonumber\\
&=\frac12\int d\bm{r}\; (\bm\nabla\bm{m}(\bm{r}))^2 
- \frac12 C_{\rm ani}\int d\bm{r}\; m_z(\bm{r})^2,
\label{eq:E}
\end{align}
which gives the effective field as 
$\bm{h}_{\rm eff}=-\delta E/\delta\bm{m}$.
The first and second terms are the interfacial and anisotropy energies,
respectively.  
When $C_{\rm ani}>0$, the $z$ component becomes dominant
since a large $m_z^2$ lowers the energy. We take $C_{\rm ani}=0.2$ in
the simulations.
The damping parameter $\alpha$ expresses the rate of energy
dissipation. 
If $\alpha=0$, the spatial average of $m_z$ as well as the energy $E$ is conserved.

Considering $m_z$ as the order parameter of this system,
we here define the characteristic length scale $L$ of a domain pattern
from the correlation function 
\begin{align}
 G(\bm{r})=\frac{1}{A}\int d^2\bm{x}\;
\langle m_z(\bm{x}+\bm{r}) m_z(\bm{x})\rangle,
\end{align}
where $A$ is the area of the system and
$\langle\cdots\rangle$ denotes an ensemble average.
The average domain size $L$ is defined by the distance where $G(r)$,
i.e., the azimuth average of $G(\bm{r})$,
first drops to zero, and thus, $G(L)=0$.

\section{\label{sec:rslt}Simulations}

\begin{figure}[tb]
 \begin{center}
  \includegraphics[width=7cm,clip]{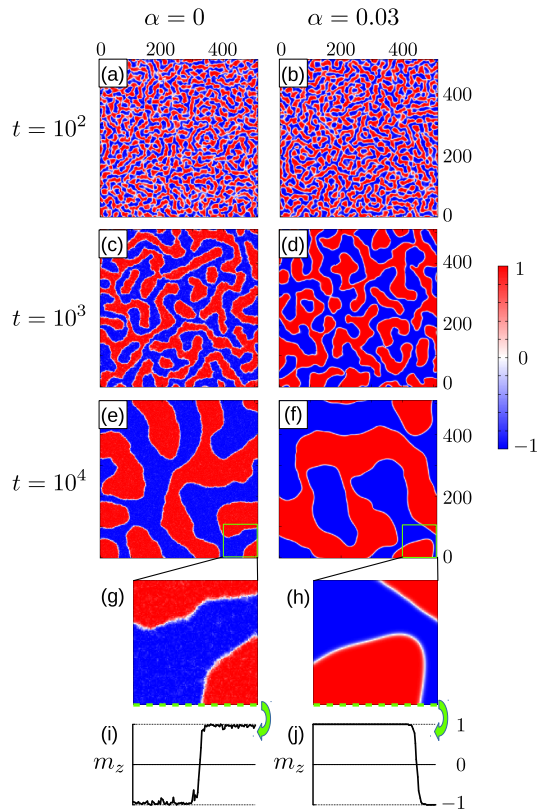}
 \end{center}
\caption{\label{fig:snap} (Color online)
Snapshots of $z$-component $m_z$ at time $t=10^2$ ((a) and (b)), 
$10^3$ ((c) and (d)), and  $10^4$ ((e) and (f)). 
Snapshots (g) and (h) are enlarged parts of (e) and (f), respectively. 
Profiles (i) and (j) of $m_z$ are taken along the bottom lines of
 snapshots (g) and (h), respectively.
Left and right columns are for the no-damping ($\alpha=0$) and damping
($\alpha=0.03$) cases, respectively.}
\end{figure}

Domain patterns appear, regardless of the damping parameter $\alpha$. 
The snapshots of the no-damping ($\alpha=0$) and
damping ($\alpha=0.03$) cases are demonstrated in the left and right
columns of Fig.~\ref{fig:snap}, respectively. 
Domain patterns at early times have no remarkable difference
between the two cases.
The characteristic length scale looks almost the same also 
at later times. However, as shown in the enlarged snapshots at late
times, difference appears especially around domain walls.
Domain walls, where $m_z\simeq 0$, are smooth in the damping
case. However, in the no-damping case, they look fuzzy.
The difference appears more clearly in profiles of $m_z$
(Figs.~\ref{fig:snap}(i) and \ref{fig:snap}(j)).
While the profile in the damping case is smooth, that in the no-damping
case is not smooth. Such an uneven profile makes domain walls look fuzzy.

\begin{figure}[tb]
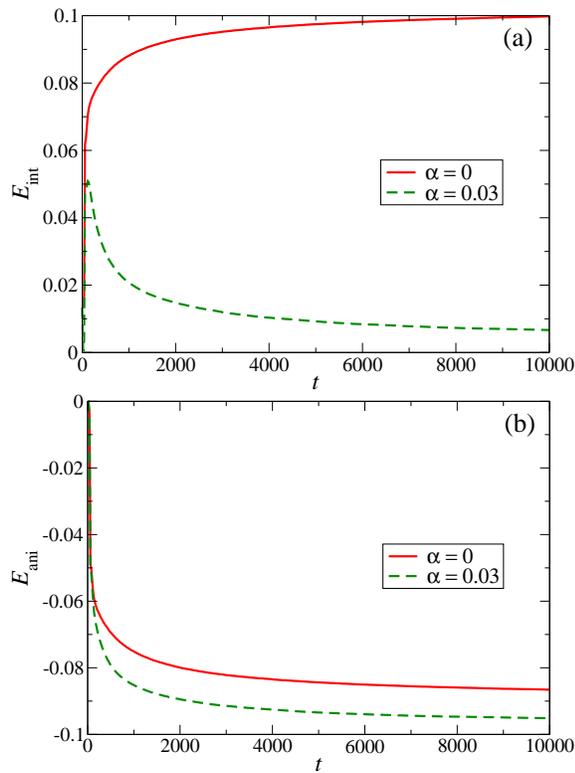

 \begin{center}
  \includegraphics[width=7.5cm,clip]{Eint.eps}
  \includegraphics[width=7.5cm,clip]{Eani.eps}
 \end{center}
\caption{\label{fig:ken} (Color online)
Time dependence of (a) the interfacial energy $E_{\rm int}$ and 
(b) the anisotropy energy $E_{\rm ani}$.
The interfacial energy increases with time 
in the no-damping case ($\alpha=0$) and
 decreases in the damping case ($\alpha=0.03$).
The anisotropy energy decreases with time in both cases.} 
\end{figure}

The difference in domain structure is closely connected with energy
dissipation, which is shown in Fig.~\ref{fig:ken}.
The interfacial energy, which is the first term of Eq.~\eqref{eq:E}, 
decays for $\alpha=0.03$ but increases for $\alpha=0$ in
Fig.~\ref{fig:ken} (a). 
In contrast, the anisotropy energy, which comes from the total of 
$m_z^2$, decreases with time for both $\alpha=0$ and
$\alpha=0.03$.
In other words, the energy dissipation relating to
the interfacial energy mainly causes the difference 
between the damping and no-damping cases. 
In the damping case, the interfacial energy decreases with time after a
shot-time increase as domain-wall structure becomes smooth. 
However, in the no-damping case, the interfacial energy increases
with time to conserve the total energy that is given by Eq.~(\ref{eq:E}). 
This corresponds to the result that the domain structure does not become
smooth in the no-damping case.
  
\begin{figure}[tb]
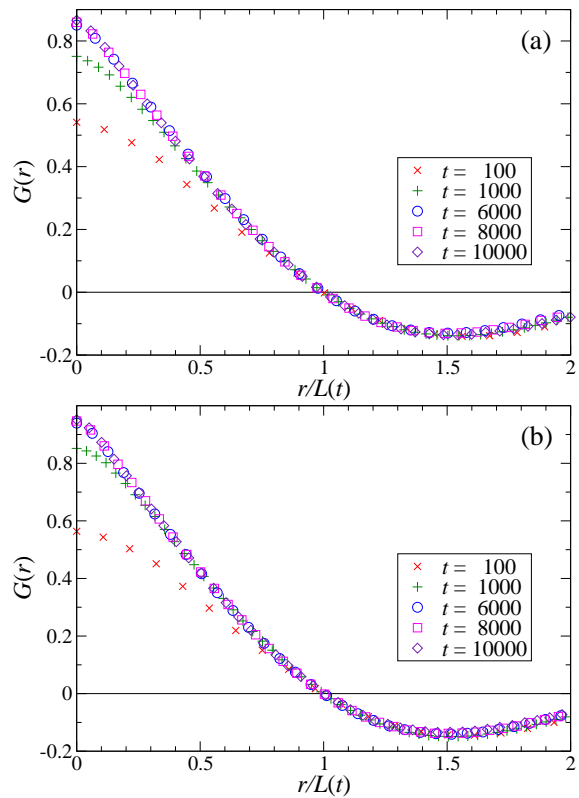

 \begin{center}
  \includegraphics[width=7.5cm,clip]{Gr0.eps}
  \includegraphics[width=7.5cm,clip]{Gr3.eps}
 \end{center}
\caption{\label{fig:Gr} (Color online)
Scaled correlation functions at different times
in (a) no-damping ($\alpha=0$) and (b) damping ($\alpha=0.03$) cases.  
The correlation functions at late times collapse to a single function,
 however, the ones at early times do not.}
\end{figure}

Before discussing growth laws, we should examine scaling laws.
Scaled correlation functions of $m_z$ at
different times are shown in Fig.~\ref{fig:Gr}.
The functions look pretty similar in both damping and no-damping cases,
which reflects the fact that the characteristic length scales in 
both cases looks almost the same in snapshots.
At late times, the correlation functions that are 
rescaled by the average domain
size $L(t)$ collapse to a single function. 
However, the scaled correlation functions at early times 
($t=100$ and $1000$) 
do not agree with the scaling function especially in the short range. 
The disagreement at early times is related with the unsaturation of
$m_z$.
How $m_z$ saturates is reflected in the time dependence of the
anisotropy energy which is shown in Fig.~\ref{fig:ken}(b).
At early times ($t\lesssim 1000$), $E_{\rm ani}$ decays rapidly. 
This implies that $m_z$ is not saturated enough in this time regime. 
The decrease in the anisotropy energy slows at late times.
In the late-time regime, $m_z$ is sufficiently saturated except for
domain walls, and the decrease in the anisotropy energy is purely caused
by domain growth. This corresponds to the scaling behavior at late
times.

\begin{figure}[tb]
 \begin{center}
  \includegraphics[width=7.5cm,clip]{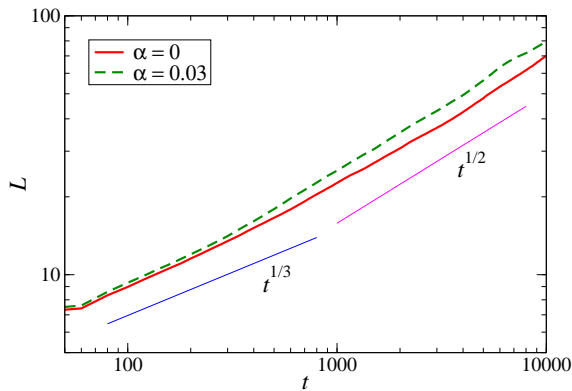}
 \end{center}
\caption{\label{fig:length} (Color online)
Time dependence of the average domain size $L$ for $\alpha=0$ and $0.03$.
In both damping and no-damping cases, domain size grows as
$L(t)\sim t^{1/2}$ at late times.
Before the scaling regime, early-time behavior looks as if
$L(t)\sim t^{1/3}$.}
\end{figure}

In Fig.~\ref{fig:length}, the average domain size $L$ is
plotted for the damping and no-damping cases.
In both cases, the average domain size grows as $L(t)\sim t^{1/2}$ at
late times, although growth exponents at early times look like $n=1/3$.  
Since scaling behavior is confirmed only at late times,
the domain growth law is considered to be $L(t)\sim t^{1/2}$ rather than
$t^{1/3}$ in this system.
In our previous work, we saw domain growth as $L(t)\sim t^{1/3}$ in a
BEC without superfluid flow~\cite{Kudo2013}, which was essentially the
same system as the present one. 
However, the time region shown in Ref.~\cite{Kudo2013}
corresponds to the early stage ($t\lesssim 1830$) in the present system. 

Although the growth exponent is supposed to be $n=1/3$ for
conserved scalar fields, 
the average domain size grows as $L(t)\sim t^{1/2}$, in our system, at
late times even in the no-damping case.
This implies that our system without damping cannot be categorized as a
model of a conserved scalar field.
Although we consider $m_z$ as the order parameter to define the
characteristic length scale, the 
LLG equation is described in terms of a vector field $\bm{m}$.

\begin{figure}[tb]
 \begin{center}
  \includegraphics[width=8cm,clip]{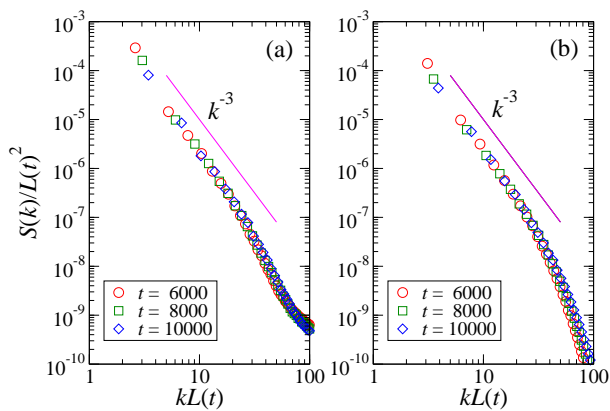}
 \end{center}
\caption{\label{fig:porod} (Color online)
Scaling plots of the structure factor scaled with $L(t)$ at different
 times in (a) no-damping ($\alpha=0$) and (b) damping ($\alpha=0.03$)
 cases. In both cases, $S(k)\sim k^{-3}$ in the high-$k$
 regime. However, they gave different tails in the ultrahigh-$k$
 regime. 
}
\end{figure}

Scaling behavior also appears in the structure factor $S(k,t)$, 
which is given by the Fourier transformation of the correlation function
$G(r)$.
According to the Porod law,
the structure factor has a power-law tail,
\begin{equation}
 S(k,t)\sim \frac{1}{L(t)k^{d+1}}, 
\label{eq:Sk}
\end{equation}
in the high-$k$ regime~\cite{Bray1994}. 
Here, $d$ is the dimension of the system.
Since $d=2$ in our system, Eq.~\eqref{eq:Sk} leads to
$S(k,t)/L(t)^2\sim [kL(t)]^{-3}$.
In Fig.~\ref{fig:porod}, $S(k,t)/L(t)^2$ is plotted as a function
of $kL(t)$.
The data at different late times collapse to one curve, and
they show $S(k)\sim k^{-3}$ in the high-$k$ regime ($kL\sim 10$) in
both the damping and no-damping cases.
In the ultrahigh-$k$ regime ($kL\sim 100$), 
tails are different between the two cases, which reflects the
difference in domain structure.
Since domain walls are fuzzy in the no-damping case, $S(k)$ remains
finite. However, in the damping case, $S(k)$ decays faster in the 
ultrahigh-$k$ regime, which is related with smooth domain walls.

\section{\label{sec:disc}Discussion}

We here have a naive question: Why does domain pattern formation occur
even in the no-damping case?
When $\alpha=0$, Eq.~\eqref{eq:LLG.0} is just the equation of the
precession of spin, and the energy $E$ as well as $m_z$ 
is conserved. 
We here discuss why similar domain patterns are formed from our initial
condition in both damping and no-damping cases. 

Using the stereographic projection of the unit sphere of spin
onto a complex plane~\cite{Laksh1984}, we rewrite Eq.~\eqref{eq:LLG.1}  
as
\begin{equation}
 \frac{\partial\omega}{\partial t} = \frac{-i+\alpha}{1+\alpha^2}\left[
\nabla^2\omega - \frac{2\omega^*(\bm\nabla \omega)^2}{1+\omega\omega^*}
- \frac{C_{\rm ani}\omega(1-\omega\omega^*)}{1+\omega\omega^*}
\right],
\label{eq:sp.1}
\end{equation}
where $\omega$ is a complex variable defined by
\begin{equation}
 \omega=\frac{m_x+im_y}{1+m_z}.
\end{equation}
Equation~\eqref{eq:sp.1} implies that the effect of the Gilbert damping
is just a rescaling of time by a complex constant~\cite{Laksh1984}.
The fixed points of Eq.~\eqref{eq:sp.1} are $|\omega|^2=1$ and
$\omega=0$. 
The linear stability analysis about these fixed points gives some clues
about domain formation.

At the fixed point $\omega=1$, $m_x=1$ and $m_y=m_z=0$, which
corresponds to the initial condition of the numerical simulation.
Substituting $\omega=1+\delta\omega$ into Eq.~\eqref{eq:sp.1}, we 
obtain linearized equations of $\delta\omega$ and $\delta\omega^*$. 
Performing Fourier expansions 
$\delta\omega=\sum_{\bm k}\delta\tilde{\omega}_{\bm k}e^{i\bm{k\cdot r}}$
and
$\delta\omega^*
=\sum_{\bm k}\delta\tilde{\omega}^*_{-\bm k}e^{i\bm{k\cdot r}}$,
we have
\begin{align}
 \frac{d}{dt}
\begin{pmatrix}
\delta\tilde{\omega}_{\bm k} \\
\delta\tilde{\omega}^*_{-\bm k}
\end{pmatrix}
=
\begin{pmatrix}
\tilde{\alpha}_1 (C_{\rm ani}-k^2) & 
\tilde{\alpha}_1 C_{\rm ani} \\
\tilde{\alpha}_2 C_{\rm ani} & 
\tilde{\alpha}_2 (C_{\rm ani}-k^2)
\end{pmatrix}
\begin{pmatrix}
\delta\tilde{\omega}_{\bm k} \\
\delta\tilde{\omega}^*_{-\bm k}
\end{pmatrix},
\label{eq:mat}
\end{align}
where $\tilde{\alpha}_1=\frac12(-i+\alpha)/(1+\alpha^2)$, 
$\tilde{\alpha}_2=\frac12(i+\alpha)/(1+\alpha^2)$, 
$\bm{k}=(k_x,k_y)$, and $k=|\bm{k}|$.
The eigenvalues of the $2\times 2$ matrix of Eq.~\eqref{eq:mat} are
\begin{align}
 \lambda(k)=\frac{\alpha}{2(1+\alpha^2)}(C_{\rm ani}-2k^2)
\pm \frac{\sqrt{4k^2(C_{\rm ani}-k^2)+\alpha^2C_{\rm ani}^2}}
{2(1+\alpha^2)}.
\label{eq:lambda}
\end{align}
Even when $\alpha=0$, $\lambda(k)$ has a positive real part for
$k < \sqrt{C_{\rm ani}}$.
Thus, the uniform pattern with $m_x=1$ is unstable, and
inhomogeneous patterns can appear.

\begin{figure}[tb]
 \begin{center}
  \includegraphics[width=7.5cm,clip]{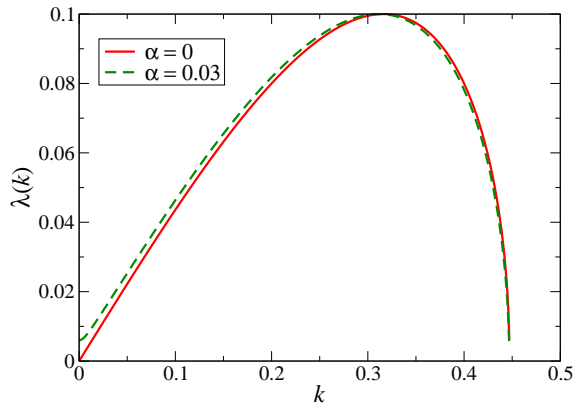}
 \end{center}
\caption{\label{fig:lambda} (Color online)
 Positive real parts of $\lambda(k)$ that is given by
 Eq.~\eqref{eq:lambda}, which has a positive real value for 
$k< \sqrt{C_{\rm ani}}$.
 The difference between $\alpha=0$ and $\alpha=0.03$ is small. 
}
\end{figure}

The positive real parts of Eq.~\eqref{eq:lambda} for
$\alpha=0$ and $\alpha=0.03$ have close values, as shown in
Fig.~\ref{fig:lambda}. This corresponds to the result that domain
formation in 
the early stage has no remarkable difference between the damping
($\alpha=0.03$) and no-damping ($\alpha=0$) cases (See
Fig.~\ref{fig:snap}).
From the view point of energy, the anisotropy energy does not necessarily
keep decaying when $\alpha=0$. For conservation of energy, it should be
also possible that both anisotropy and interfacial energies change only a
little. 
Because of the instability of the initial state, $m_z$ grows, and thus,
the anisotropy energy decreases.

The initial condition, which is given as spins aligned in one
direction with some noises in the $x$-$y$ plane, 
is the key to observe domain pattern
formation in the no-damping case.
Actually, if spins have totally random directions, 
no large domains are formed in the no-damping case,
although domains are formed in damping cases
($\alpha>0$) from such an initial state. 

When $\omega=0$, $m_x=m_y=0$ and $m_z=1$, which is also one of the fixed 
points. 
Substituting $\omega=0+\delta\omega$ into Eq.~\eqref{eq:sp.1} and
performing Fourier expansions, we have the linearized equation of
$\delta\tilde{\omega}_{\bm k}$,
\begin{align}
 \frac{d}{dt}\delta\tilde{\omega}_{\bm k}=
\frac{i-\alpha}{1+\alpha^2}(k^2+C_{\rm ani})\delta\tilde{\omega}_{\bm k}.
\end{align}
This implies that the fixed point is stable for $\alpha>0$ and 
neutrally stable for $\alpha=0$.
Although $m_z=-1$ corresponds to $\omega\to\infty$, the same stability is
expected for $m_z=-1$ by symmetry.

Since the initial condition is unstable, the $z$-component of spin
grows. Moreover, linear instability is similar for $\alpha=0$ and
$\alpha=0.03$. 
Since $m_z=\pm1$ are not unstable, 
$m_z$ can keep its value at around $m_z=\pm1$.
This is why similar domain patters are formed in both damping and
no-damping cases. The main difference between the two cases is that
$m_z=\pm1$ are attracting for $\alpha>0$ and neutrally stable for
$\alpha =0$. 
Since $m_z=\pm1$ are stable and attracting in the damping case,
homogeneous domains with $m_z=\pm1$ are preferable, which leads to a
smooth profile of $m_z$ such as Fig.~\ref{fig:snap}(j).
In the damping case, $m_z=\pm1$ are neutrally stable (not
attracting) fixed points, which does not necessarily make domains
smooth.  

\section{\label{sec:conc}Conclusions}

We have investigated the domain formation in
2D vector fields with an easy-axis anisotropy, using
the LLG equation.
When the initial configuration is given as almost uniform spins aligned
in an in-plane direction, similar domain patterns appear in 
the damping ($\alpha\neq 0$) and no-damping ($\alpha=0$) cases.
The average domain size grows as $L(t)\sim t^{1/2}$ in late times
which are in a scaling regime. 
The damping gives no remarkable effects on 
domain growth and large-scale properties of domain pattern.
In contrast, small-scale structures are different
between the two cases, which is shown quantitatively in the structure
factor. This difference is induced by the reduction of the interfacial
energy due to the damping.
It should be noted that the result and analysis especially in the
no-damping case are valid for a limited initial condition.
Although domains grow in a damping case even from spins with totally random
directions, domain growth cannot occur
from such a random configuration in the no-damping case. 

\begin{acknowledgments}
This work was supported by MEXT KAKENHI (No.~26103514, ``Fluctuation \&
 Structure'').
\end{acknowledgments}


\providecommand{\noopsort}[1]{}\providecommand{\singleletter}[1]{#1}%

\end{document}